\documentclass[prl,nofootinbib,superscriptaddress]{revtex4}
\usepackage[a4paper,left=1.5cm,right=1.5cm,top=3cm,bottom=3cm]{geometry}
\usepackage[toc]{appendix} 

\usepackage{amssymb,amsmath,amsfonts}
\usepackage[dvipsnames]{xcolor}
\usepackage{graphicx}
\usepackage{longtable}
\usepackage{verbatim}
\usepackage{color}
\usepackage[mathscr]{euscript}
\usepackage{framed}
\usepackage{amsmath}
\usepackage{mdframed}
\usepackage{amsmath}
\usepackage{amsfonts}
\usepackage{mathrsfs}
\usepackage{amssymb}
\usepackage{mathrsfs}  
\newcommand{\arXiv}[2]{\href{http://arxiv.org/pdf/#1}{{\tt [#2/#1]}}}
\newcommand{\arXivold}[1]{\href{http://arxiv.org/pdf/#1}{{\tt [#1]}}}

\newcommand{\nc}{\newcommand}
\nc{\ba}{\begin{eqnarray}}
\nc{\ea}{\end{eqnarray}}

\newcommand{\calR}{{\cal{R}}}
\newcommand{\calP}{{\cal{P}}}

\def\bfp{{\bf p}}
\def\bfk{{\bf k}}
\def\bfq{{\bf q}}

\usepackage{color}
\usepackage{hyperref}
\hypersetup{colorlinks, citecolor=bluscuro, linkcolor=black, urlcolor=bluscuro}
\definecolor{rossos}{cmyk}{0,1,1,0.55}
\definecolor{bluscuro}{rgb}{0.15, 0.2, .85}
\definecolor{bluchiaro}{cmyk}{1,.3,0.,0.1}

\graphicspath{{./Figures/}}

\def\ltsim{\lower3pt\hbox{$\, \buildrel < \over \sim \, $}}  
\def\gtsim{\lower3pt\hbox{$\, \buildrel > \over \sim \, $}}  
\newcommand{\be}{\begin{equation}}  
\newcommand{\ee}{\end{equation}}  
\def\ga{\mathrel{\raise.3ex\hbox{$>$\kern-.75em\lower1ex\hbox{$\sim$}}}}  
\def\la{\mathrel{\raise.3ex\hbox{$<$\kern-.75em\lower1ex\hbox{$\sim$}}}}  

\makeatletter   
\@addtoreset{equation}{section}   
\makeatother   
   
\numberwithin{equation}{section}

\newcommand{\doi}[2]{\href{https://dx.doi.org/#1}{#2}}

\usepackage[dvipsnames]{xcolor}
\usepackage{hyperref}
\hypersetup{colorlinks, citecolor=ForestGreen, linkcolor=BrickRed, urlcolor=NavyBlue}
\makeatletter   
\@addtoreset{equation}{section}   
\makeatother

\usepackage[normalem]{ulem}

\begin{document}  
  
\baselineskip=18pt   
\begin{titlepage}  
\begin{center}  
\vspace{0.5cm}  
  
\Large {\bf Primordial Black Holes and Loops in  Single-Field Inflation}

\vspace*{20mm}  
\normalsize  

{\large\bf Hassan Firouzjahi$^{1}$ and 
 Antonio ~Riotto$^{2,3}$  }

\smallskip   
\medskip   
\vskip 0.5cm

\it{\small $^1$School of Astronomy, Institute for Research in Fundamental Sciences (IPM)\\
P. O. Box 19395-5531, Tehran, Iran}

\it{\small $^2$ Department of Theoretical Physics and Gravitational Wave Science Center,  \\
			24 quai E. Ansermet, CH-1211 Geneva 4, Switzerland}

  \it{\small $^3$ LAPTh, CNRS, USMB, F-74940 Annecy, France}
  \end{center}
  
\vskip0.6in

\centerline{\bf Abstract}
\vskip 0.5cm  
\noindent
Using  the $\delta N$ formalism we calculate  the one-loop correction to the large-scale power spectrum of the curvature perturbation in the standard scenario where primordial black holes are formed  in the early universe thanks to  a phase of ultra-slow-roll in single-field inflation. We explicitly show that one-loop  corrections are negligible when the transition from the 
ultra-slow-roll to the slow-roll phase is smooth.  We conclude that the PBH formation  scenario through a ultra-slow-roll phase is viable.

\vspace*{2mm}   

\end{titlepage} 

\section{I. Introduction}  \label{sec:intro}
\setcounter{section}{1}
\noindent
 A  standard  scenario where   Primordial Black Holes (PBHs) are generated  in the early universe  is  by enhancing   the curvature perturbation  at short scales  \cite{s1,s2,s3} during an Ultra Slow Roll (USR)  period when  the inflaton potential  $V(\phi)$ becomes sufficiently  flat. 
The $\delta N$ formalism is a powerful approach to calculate the curvature perturbation at the end of inflation \cite{deltaN1,deltaN2, Abolhasani:2019cqw}. Based on the  separate universe approach, it is able to capture the super-Hubble evolutions of the perturbations beyond the linear order, thus capturing the effects of  non-Gaussianity and providing a useful set-up to calculate the loop corrections \cite{deltaNloop}.
 
In this  note we use the $\delta N$ formalism to calculate the one-loop corrections to the very large scale (i.e. CMB scale) power spectrum of the curvature perturbation induced by short-scale curvature modes. The latter modes, as the seeds of PBHs formation,  are enhanced during the USR phase which is preceded and followed by the  standard Slow-Roll (SR) phases. The issue of the impact of the one-loop correction from the short modes onto the large-scale power spectrum has been actively debated in the recent literature (see for instance Refs. \cite{Kristiano:2022maq, riotto1, Choudhury:2023vuj, yokoyamanew, riotto2,f,m}). 

We show explicitly that loop corrections are suppressed in the case in which the transition between the USR and the SR phase is smooth, as originally advocated in Refs. \cite{riotto1,riotto2}.

The paper is organised as follows. In section II we discuss the $\delta N$ formalism, loop corrections are calculated in section III and conclusions in section IV.


\section{II. Three-phase  $\delta N$ analysis} 
\setcounter{section}{2}
The crucial starting point of our calculation is to take into account the SR phase preceding the USR phase. As we shall see, this turns out to be  essential to capture the dependence of the long CMB mode onto the short modes. 

We consider  therefore a three-step model of inflation:  SR $\rightarrow$ USR $\rightarrow$ SR. In our convention the CMB scale modes are denoted by 
$\bfp$ while those of small scales running in the loop by $\bfq$. Our goal is to extend the original $\delta N$ analysis in Refs. \cite{Namjoo:2012aa, Chen:2013eea} and \cite{Cai:2018dkf} to the case where the mode of interest $\bfp$ leaves the horizon during the first SR phase. 

An important step is that we should keep track of the decaying mode in the first SR phase. During the first SR phase the decaying mode falls off exponentially. However, during the intermediate USR phase, this decaying mode is growing and it is behind the evolution of curvature perturbations on superhorizon scales. Therefore, for a consistent treatment we have to glue the decaying modes at the transition from the first SR phase to the USR phase. For this purpose, we have to solve the second order Klein-Gordon equation in the first SR phase to keep track of the decaying mode. In the spirit, this is similar to the method employed in Ref. \cite{Cai:2018dkf} who studied a second order differential equation in the final SR phase to keep track of the evolution of the mode perturbations.

\subsection{First SR phase}
\noindent
With the above discussion in mind, let us start with the first SR phase. 
We assume the field starts at the initial
position $\phi_i$ where the CMB scale mode $\bfp$ leaves the horizon. The first SR phase ends at $t=t_s$ when the USR phase starts. The USR phase itself is extended in the period $t_s \leq t \leq t_e$. It is also important for our $\delta N$ formalism that the positions of the start and the end of USR phase, 
$\phi_s\equiv  \phi(t_s)$ and $\phi_e \equiv  \phi(t_e)$, are fixed so they are not varied when performing  the variation $\delta N$.

To keep track of the evolution of mode function during the first SR phase, similar to  Ref. \cite{Cai:2018dkf}, we expand the  potential near the point $\phi=\phi_s$ as follows
\ba
V(\phi) = V(\phi_s) + \sqrt{2 \epsilon_s} V(\phi_s)  (\phi -\phi_s) + \frac{\eta_s}{2} V(\phi_s) (\phi -\phi_s)^2 + \cdots \, .
\ea
Here  the  SR parameters $\epsilon_s$ and $\eta_s$ are 
defined in terms if the derivatives of the potential,  
\ba
\epsilon_s= \frac{1}{2} \left( \frac{V'(\phi_s)}{V(\phi_s)} \right)^2 \, , \quad \quad
\eta_s=  \frac{V''(\phi_s)}{V(\phi_s)}  \, ,
\ea
where a prime denote the derivative with respect to the field $\phi$ and we work with the convention that the reduced Planck mass is set to unity, $M_P=1$.  
Note that to leading orders in the SR  parameters, 
\ba
\label{epsilon-app}
\epsilon_s \simeq \frac{\dot \phi_s^2}{2 M_P^2 H^2} \simeq -\left.\frac{\dot H}{H^2}\right|_{t_s} \, ,
\ea
in which a dot represents the derivative with respect to cosmic time and 
$H$ is the Hubble expansion rate during inflation. 
To prevent the confusion, we define the other SR parameter associated to the time derivative of $\epsilon$ as $ \eta \equiv \dot \epsilon/(H \epsilon)$. During the USR phase $ \eta\simeq -6$ which should not be mistaken with $\eta_s \ll 1$. 

Using the number of e-olds $N$ as the clock via $dN = H d t$, the evolution of the field during the first SR phase is given by
\ba
\label{ode1}
 \frac{d^2 \phi}{ d N^2} + 3 \frac{d \phi}{d N} + 3  \sqrt{2 \epsilon_s} + 3 \eta_s (\phi - \phi_s) 
 \simeq 0 \, ,
 \quad \quad 3 H^2 \simeq V(\phi_s) \, ,
 \ea
whose solution is 
\ba
\label{sol1}
\phi(N)= C_1 e^{-\alpha N} + C_2 e^{-\beta N} + \left(  \phi_s -\frac{\sqrt{2 \epsilon_s}}{\eta_s} \right) \, ,
\ea
in which $C_1$ and $C_2$ are two constants of integration and,
\ba
\label{alpha-beta}
\alpha  &\equiv& \frac{1}{2} \left( -3 + \sqrt{9 - 12 \eta_s} \right) \simeq -\eta_s \, , 
\nonumber\\
\beta  &\equiv& \frac{1}{2} \left( -3 - \sqrt{9 - 12 \eta_s} \right) \simeq -3 + \eta_s \, .
\ea
The decaying mode is controlled mainly by the evolution of the momentum $\pi\equiv d \phi/ d N$. Therefore, it is better to use the phase space variables $(\phi, \pi)$ and then express the constants $C_1$ and $C_2$ in terms of $(\phi(N), \pi(N))$. This way, we can use the values of $C_1$ and $C_2$ to match the solution $\big(\phi(N), \pi(N) \big)$
with the initial  conditions $(\phi_i, \pi_i)$ to their corresponding values $(\phi_s, \pi_s)$ at the time  of the start of USR phase, $t=t_s$. 

Solving $C_1$ and $C_2$ in terms of  $(\phi(N), \pi(N))$, we obtain
\ba
\label{C-vals}
C_1 &=& e^{\alpha N}  \, \frac{\pi+ \beta \widetilde \phi}{\beta- \alpha} \, , \nonumber\\
C_2 &=& - e^{\beta N} \,  \frac{\pi+ \alpha \widetilde \phi}{\beta- \alpha} \, .
\ea
Here, for convenience, we have defined
\ba
\label{tildephi}
\widetilde \phi \equiv  \phi - \phi_s + \frac{\sqrt{2 \epsilon_s}}{\eta_s} \, .
\ea
The above expressions for $(C_1, C_2)$ are valid for anytime during $t \leq t_s$. Therefore, we can relate $(\phi_s, \pi_s)$ to $\left(\phi(N), \pi(N) \right)$ for any time during the first SR phase and in particular to the initial values $(\phi_i, \pi_i)$. 
More specifically, we obtain the following relation between $(\phi_s, \pi_s)$, $(\phi_i, \pi_i)$, $N_i$ and $N_s$:
\ba
\label{eqN1}
e^{\alpha ( N_s- N_i )} = \frac{\pi+ \beta \widetilde \phi_i}{\pi_s+ \beta \widetilde \phi_s} \, ,
\ea
and
\ba
\label{eqN2}
e^{\beta ( N_s- N_i )} = \frac{\pi+ \alpha \widetilde \phi_i}{\pi_s+ \alpha \widetilde \phi_s}\, .
\ea
Note that , as in Ref. \cite{Cai:2018dkf}, we have choses the convention that $N=0$ at the end of USR phase, so during the first SR phase and during USR phase $N<0$. Also $N_i$ is the initial value of $N$ while the time of start of USR is denoted by $N_s$, such that  $N_i, N_s <0$.  For the CMB scale modes $N_i \sim -30$ assuming that  the USR phase happens somewhere in the middle of inflation to generate the right mass scale for PBHs formation. Also, typically, $|N_s| \lesssim$ few such that
$|N_i| \gg | N_s|$. 

From Eq. (\ref{eqN2}), we see that $\pi_s+ \alpha \widetilde \phi_s$ falls off exponentially to zero  like $e^{- 3 ( N_s- N_i )} 0$. This is the hallmark of the decaying mode in the first SR phase. Therefore, to a very good accuracy, and using Eq. (\ref{epsilon-app}), we have $\pi_s \simeq - \sqrt{2 \epsilon_s}$. This approximation can be used safely in Eq. (\ref{eqN1}) to solve for $N_s- N_i$, yielding
\ba
\label{Ns-N}
N_s- N_i &=&\frac{1}{\alpha} \ln \left(\frac{\pi_i + \beta \widetilde \phi_i}{\pi_s + \beta \widetilde {\phi_s}} \right)  \simeq \frac{1}{\eta_s}  \ln \left(\frac{\pi_i + \beta \widetilde \phi}{ -3 \frac{\sqrt{2 \epsilon_s}}{\eta_s}} \right) \, \nonumber\\
&\simeq & \frac{1}{\eta_s}  \ln \left( 1 + \eta_s \frac{\phi_i - \phi_s}{\sqrt{2 \epsilon_s}} 
-\frac{\eta_s}{3 \sqrt{2 \epsilon_s}} \left( \pi_i + \sqrt{2 \epsilon_s} \right) \right) \, .
\ea
Note that for practical purpose, we can ignore the last term in the big bracket above since $\pi \simeq - \sqrt{2 \epsilon_s}$ to very good accuracy as described above. 
Now, plugging this expression for $N_s- N_i $ in Eq. (\ref{eqN2}), we can relate $\pi_s$ to the initial values $(\phi_i, \pi_i)$ as follows
\ba
\label{pis}
\pi_s \simeq - \sqrt{2 \epsilon_s} +  (\pi_i + \eta_s \widetilde \phi_i) 
\left( 1 + \eta_s \frac{\phi_i - \phi_s}{\sqrt{2 \epsilon_s}}  \right)^{-\frac{3}{\eta}}
\, .
\ea
Since the last term falls off exponentially for our setup in which $N_s- N_i \sim 30$ or so, then we can neglect for all practical purposes the contributions of $\delta \pi_s$ in $\delta N$.

\subsection{The USR phase}
\noindent
During the USR phase we have
\ba
\ddot \phi + 3 H \dot \phi =0 \, .
\ea
Similar to the first phase we solve for $(\phi(N), \pi(N))$ as follows
\ba
\pi(N) = D_1 e^{-3 N}  \, ,  \quad \quad  \phi(N) = -\frac{\pi(N)}{3} + D_2 \, ,
\ea
in which $D_1$ and $D_2$ are two constants of integrations. 
Eliminating the constants $D_1$ and $D_2$, we obtain the following relations,
\ba
\label{pis-pie}
\pi_e = \pi_s e^{3 N_s},    \quad \quad \pi_e = \pi_s + 3 ( \phi_s - \phi_e) \, ,
\ea
in which $\pi_e$ is the value of $\pi$ at the time  $t_e$. As mentioned previously, it is important to note that $\phi_s$ and $\phi_e$ are fixed while $\pi_s$ and $\pi_e$ can vary as functions of the initial values in phase space $(\phi_i, \pi_i)$.  In particular, from the second equation above we find $\delta \pi_e = -\delta \pi_s$.

The relations in Eq. (\ref{pis-pie}) can then be combined, to yield the following expression for $N_s$:
\ba
\label{Ns-eq}
N_s= \frac{1}{3} \ln \left(\frac{\pi_e}{\pi_s} \right) = \frac{1}{3} \ln \Big[1+ \frac{3 (\phi_s - \phi_e)}{\pi_s} \Big]   \, .
\ea
A crucial difference compared to the setup of  Ref. \cite{Cai:2018dkf} is that the contribution of $N_s$ in total $\delta N$ is quite negligible. This is because since $\phi_s$ and $\phi_e$ are fixed,  the only way in  which $N_s$ can contribute to $\delta N$ is  through $\pi_s$. However, $\pi_s$ is exponentially close to $-\sqrt{2 \epsilon_s}$ as given in Eq. (\ref{pis}),  so basically we can neglect the contribution of $\delta \pi_s$ in $\delta N$  in our calculation.

Another important point to notice is that as $\pi$ falls off exponentially during the USR phase, the SR parameter $\epsilon$ falls off exponentially as well. More specifically, at the end of USR phase we have $\epsilon_e= \epsilon_s e^{-6 |N_s|}$.
As a result, for the modes which leave the horizon during the USR phase  the  curvature perturbation grows as   $\calR \propto a^3(t) \propto e^{-3 N}$ in which $a(t)$ is the FLRW scale factor.

\subsection{The final  SR phase}
\noindent
Finally, the evolution of the field in the final SR phase and the contribution to $\delta N$ is the same as in Ref. \cite{Cai:2018dkf}. This time, we expand the potential around 
$\phi_e$  as follows 
\ba
V(\phi) = V(\phi_e) + \sqrt{2 \epsilon_V} V(\phi_e)  (\phi -\phi_e) + \frac{\eta_V}{2} V(\phi_e) (\phi -\phi_e)^2 + \cdots \, .
\ea
Here $2\epsilon_V \equiv \left(V'(\phi_e)/V(\phi_e) \right)^2$ and $\eta_V\equiv V''(\phi_e)/V(\phi_e)$ are the  SR parameters defined at the point $\phi_e$. 
 As the potential supports a SR phase after the USR period and the system approaches its attractor SR phase shortly afterwards, we can assume  $ \epsilon_V$ and $ \eta_V$ to be the corresponding SR parameters during the final SR phase. We also allow for the possibilities that $\epsilon_s \neq \epsilon_V$ and $\eta_s \neq \eta_V$.

The evolution of the field is then given by
\ba
 \frac{d^2 \phi}{ d N^2} + 3 \frac{d \phi}{d N} + 3  \sqrt{2 \epsilon_V} + 3 \eta_V (\phi - \phi_e) 
 \simeq 0 \, ,
 \quad \quad 3  H^2 \simeq V(\phi_e) \, .
 \ea
The value of $N$ at the end of attractor phase in the final SR regime, 
$N_f$, has been obtained in  Ref. \cite{Cai:2018dkf} 
\ba
\label{Nf}
N_f= \frac{1}{\eta_V} \ln\Big[-2 \eta_V \pi_e - 6\sqrt{2 \epsilon_V}  M_P  \Big]\, ,
\ea
with the  important  point that the dependence of $N_f$ to the initial values 
$(\phi_i, \pi_i)$ comes only through $\pi_e$. However, as mentioned previously, with fixed values of $\phi_s$ and $\phi_e$, from Eq. (\ref{Ns-eq}) we see that 
$\delta \pi_e=-\delta \pi_s$. Since the dependence of $\pi_s$ to the initial values 
$(\phi_i, \pi_i)$ are exponentially suppressed, we conclude that the contribution of the final duration of inflation $N_f$ to $\delta N$ associated to initial conditions 
$(\phi_i, \pi_i)$ are exponentially suppressed as well. 

As mentioned before, we allow for the possibility that $\epsilon_s \neq \epsilon_V$
and $\eta_s \neq \eta_V$. However, in a natural situation where  the shapes of the potential before and after the USR phase are similar, we typically expect that $\epsilon_s \sim \epsilon_V$ and $\eta_s \sim \eta_V$.
However, it is important to realize that there can be a large hierarchy between $\epsilon_e=\pi^2_e/2$ and $\epsilon^2_V$. This hierarchy is a measure of the sharpness of the transition. Following the convention of \cite{Cai:2018dkf}  we define the sharpness parameter  $h$ as follows
\ba
h\equiv 6 \frac{\sqrt{2 \epsilon_V}}{\pi_e}=-6\sqrt{\frac{\epsilon_V}{\epsilon_e}}\, .
\ea 
For a sharp transition we have $h\leq -6$ while for a smooth transition we have 
$h \rightarrow 0$. 

\subsection{The curvature perturbation from the $\delta N$ formalism}
\noindent
At this stage, we are ready to calculate the curvature perturbation at non-linear order through the $\delta N$ formalism. The total number of e-folds is given by
$N(\phi_i, \pi_i)\equiv  (N_f- N_i)$. Using Eq. (\ref{Ns-N}) for $N_i$ with $N_s$ given in Eq. (\ref{Ns-eq}) we obtain 
\ba
\label{N-eq-final}
N(\phi_i, \pi_i)\simeq  \frac{1}{\eta_s} \ln \Big[ 1+ \frac{\eta_s}{\sqrt{2 \epsilon_s}  M_P} \left( \phi_i - \phi_s \right)  \Big]- \frac{1}{3} \ln \Big[1+ \frac{3 (\phi_s - \phi_e)}{\pi_s(\phi_i, \pi_i)} \Big] 
 + \frac{1}{\eta_V} \ln\Big[ -2 \eta_V \pi_e - 6\sqrt{2 \epsilon_V}  M_P  \Big] \, .
\ea
There are a number of important differences compared to the result of  Ref. \cite{Cai:2018dkf}.
The first difference is that we have the first term above coming from the first SR phase. Second, the variation is with respect to $\delta \phi_i$ which leaves the horizon at the initial flat hypersurface in the first SR phase while $\phi_s$ and $\phi_e$ are held fixed.
Also note that the dependence to $\pi_i$ comes only via $\pi_s$ which is exponentially suppressed. 

As discussed before, the leading contribution to 
$\delta N$ comes from the first term in Eq. (\ref{N-eq-final}) which is the usual contribution from  single field SR inflation. This is consistent with the intuition that the long CMB scale modes which have left the horizon during the early SR phase 
should be largely unaffected during the subsequent evolution of the inflationary background.

Having the  explicit function $N(\phi_i)$ we can proceed with the $\delta N$ analysis and expand around the background value $\phi_i$ as follows 
\ba
\label{deltaN}
\delta N = N' \delta \phi + \frac{N''}{2} \delta \phi^2 + \frac{N'''}{3!} \delta \phi^3 +\cdots \, ,
\ea
where by $\delta \phi$ we actually mean $\delta \phi_i$ calculated at the time of horizon crossing on a flat hypersurface in the first SR phase. Also a prime denotes the derivative with respect to $\phi$.

The curvature  perturbation is $\calR = -\delta N$ and correspondingly, 
the power spectrum $P_\calR$ at the linear order is 
\ba
\label{PR}
P_\calR(\bfp) =  N'^2 P_{\delta \phi} = N'^2  \left( \frac{H^2}{2 p^3} \right)  \, .
\ea
On the other hand, from Eq. (\ref{N-eq-final}) we see that to the leading order 
\ba
\label{N-prime}
N'\simeq \frac{1}{\sqrt{2 \epsilon_s}}  \Big[ 1+ \frac{\eta_s}{\sqrt{2 \epsilon_s}  M_P} \left( \phi_i - \phi_s \right)  \Big]^{-1} \simeq  \frac{1}{\sqrt{2 \epsilon_s}} + {\cal O}(\eta_s) \, .
\ea
The contributions of the second and third terms in Eq. (\ref{N-eq-final}) contain the additional factors $e^{-3 (N_s- N_i)}$ and $e^{-3  N_i} e^{-\eta_V N_f}$, respectively. However, in our setup in which $N_i \sim -30$ and $N_s \sim \mathrm{few}$ these contributions   are exponentially suppressed compared to the contribution from the first term as given in Eq. (\ref{N-prime}). 

Defining the dimensionless power spectrum via 
$\calP_\calR = (p^3/2 \pi^2) P_\calR(p)$, the power spectrum of CMB scale modes which leave the horizon during the first phase of inflation is given by
\ba
\label{power}
\calP_\calR \simeq N'^2 \calP_{\delta \phi} \simeq \frac{H^2}{8 \pi^2  \epsilon_s}
+ {\cal O}\left(e^{-3 (N_s- N_i)} \right) \, .
\ea
At this stage, it is also instructive to look at the bispectrum   and the $f_{NL}$ parameter. Defining the bispectrum as 
$B_\calR(\bfk_1, \bfk_2, \bfk_3)  \equiv (2 \pi)^3 \delta^3 (\bfk_1+ \bfk_2+ \bfk_3) 
 \langle \calR(\bfk_1) \calR(\bfk_2)\calR(\bfk_3)\rangle $, the $f_{NL}$ parameter is related to $B_\calR(\bfk_1, \bfk_2, \bfk_3) $ via \cite{deltaNloop}
 \ba
 B_\calR(\bfk_1, \bfk_2, \bfk_3)   = \frac{6}{5} f_{NL} 
 \big[ P_\calR(k_1)  P_\calR(k_2) +  P_\calR(k_1)  P_\calR(k_3)+  P_\calR(k_2)  P_\calR(k_3) \big] \, .
 \ea
With the expansion of $\delta N$ as given in Eq. (\ref{deltaN}) we obtain \cite{deltaNloop}
\ba
\label{fNL}
f_{NL}= \frac{5}{6} \frac{N''}{N'^2} \, .
\ea
With the logic explained above, the dominant contribution in $N''$ comes from the first SR phase of inflation with 
\ba
N'' \simeq \frac{-\eta_s}{2 \epsilon_s}  \Big[ 1+ \frac{\eta_s}{\sqrt{2 \epsilon_s}  M_P} \left( \phi_i - \phi_s \right)  \Big]^{-2} \, ,
\ea
yielding
\ba
f_{NL} \simeq -\frac{5}{6} \eta_s +  {\cal O}\left(e^{-3 (N_s- N_i)} \right) \, .
\ea
The above results for the power spectrum and bispectrum confirm the physical expectation that at the tree level the CMB scale modes are largely unaffected by the small scale modes while the latter modes  experiences a growth during the intermediate USR phase.

Also, for later reference we note that 
\ba
\label{N3-prime}
N''' \simeq \frac{2 \eta_s^2}{( 2 \epsilon_s)^{\frac{3}{2}}} \Big[ 1+ \frac{\eta_s}{\sqrt{2 \epsilon_s}  M_P} \left( \phi_i - \phi_s \right)  \Big]^{-2}
\ea
and correspondingly
\ba
\frac{N'''}{N'^3} \simeq 2 \eta_s^2 \, ,  \quad \quad
\frac{N'''}{N'} \simeq  \frac{\eta_s^2}{ \epsilon_s} \, .
\ea


\section{III. The large scale   Power Spectrum at one-loop}
\setcounter{section}{3}
\noindent
To calculate the loop corrections, let us look at the Fourier component of $\delta N$ as given in Eq. (\ref{deltaN})
\ba
\label{Fourier1}
\delta N_\bfp =  N' \delta \phi_\bfp + \frac{N''}{2} \int \frac{d^3 \bfq}{( 2 \pi)^3} 
\delta \phi_\bfq \delta \phi_{\bfp- \bfq}
+ \frac{N'''}{3!} \int \frac{d^3 \bfq_1}{( 2 \pi)^3} \frac{d^3 \bfq_2}{( 2 \pi)^3} 
\delta \phi_{\bfq_1}  \delta \phi_{\bfq_2} \delta \phi_{\bfp- \bfq_1- \bfq_2}\, .
\ea
Correspondingly, the two-point correlation function is given by the following series
\ba
\label{power-total}
\langle \delta N_{\bfp_1} \delta N_{\bfp_2} \rangle  \equiv (2 \pi)^3 \delta^3 (\bfp_1+ \bfp_2)  \left( A_1 + A_2 + A_3+\cdots \right) \, ,
\ea
in which $A_i$ are given as follow. In constructing the terms $A_i$, we note that 
since $p \ll q$, the leading terms are only those which contain at least one component  $\delta \phi_{\bfp_1}$ multiplied by convolution integrals over $d^3 \bfq_j$. For example, a term like $ \int d^3 \bfq_1 \int d^3 \bfq_2 \langle \delta \phi_{\bfq_1} \delta \phi_{\bfp_1- \bfq_1} 
\phi_{\bfq_2} \delta \phi_{\bfp_1- \bfq_2}\rangle$ is subleading as this term does not lead to the extra factor $p_1^{-3}$ as required  for the scaling of power spectrum.

Starting  with $A_1$ we have 
\ba
\label{A1}
A_1 \equiv N'^2 \langle \delta \phi_{\bfp} \delta \phi_{-\bfp} \rangle\, .
\ea
In particular, note that at the tree-level in which $\langle \delta \phi_{\bfp} \delta \phi_{-\bfp} \rangle = (H^2/2 p^3) $
the term $A_1$ is the power spectrum as given in Eq. (\ref{PR}). However, there are loop corrections in $A_1$ as well. This is because in $\delta N$ formalism, $\langle \delta \phi_{\bfp} \delta \phi_{-\bfp} \rangle$ should be calculated on a flat hypersurface as an initial input value. Suppose we go to the flat gauge in which 
$\calR = -(H/\dot \phi) \delta \phi$. Expanding the potential around its background value, and going to decoupling limit where the metric takes its unperturbed FLRW form, we have the  interactions $V''' \delta \phi^3$ and $V'''' \delta \phi^4$. 

In the SR setup or in a case with a smooth transition from a USR phase to a SR phase, both $V'''$ and  $V''''$ are SR suppressed. Therefore, the corresponding interactions 
$V''' \delta \phi^3$ and $V'''' \delta \phi^4$ contains additional factors of SR parameters and can be neglected to leading order. However, in a sharp transition 
in which $| h| \gg 1$, there can be large loop corrections from these interactions. 
The analysis  for the loop corrections in $A_1$ for the limit of sharp transition were  performed specifically in \cite{f}  which we do not repeat here. We comment that the large interactions from
$V'''$ and $V''''$ contain the derivatives of the low-roll parameter $\tilde \eta = \dot \epsilon/H \epsilon$. So in a sharp transition from a USR phase to a SR phase, $\dot {\tilde \eta} \sim  \delta (t- t_e)$ which can induce large loop corrections as in Ref. \cite{Kristiano:2022maq}. 

However, as just mentioned and already advocated in Refs. \cite{riotto1,riotto2}, in the limit  of a smooth transition, the interactions with the vertices $V'''$ and  $V''''$ come with extra factor of SR parameter and their contributions are suppressed compared to dangerous loop corrections 
calculated in the limit of sharp transition  of Refs. \cite{Kristiano:2022maq} and \cite{f}.

The term $A_2$ is given by
\ba
\label{A2}
A_2\equiv N' N''  \int \frac{d^3 \bfq}{( 2 \pi)^3} 
\big \langle    \delta \phi_{\bfp_1}  \delta \phi_\bfq \delta \phi_{-\bfp_1- \bfq}
\big\rangle \, . 
\ea
It depends on the intrinsic bispectrum of the fluctuations $\delta\phi$ and 
 can yield a non-zero value if there is intrinsic non-Gaussianity in the system.
This happens when we have $\delta \phi^3$ interaction in the model, that is an interaction term in the Lagrangian of the form 
\be
\label{L3}
{\cal L}_3\sim  a^3V'''\delta\phi^3\, .
\ee
For a smooth USR to SR transition, as long as we have a
SR potential, that is $V'''$ small, the non-Gaussianities would be always small. 
Of course, if the potential is not smooth around the transition between the USR and the SR phase,  this  may yield large $V'''$ and loop corrections may be large. 

To calculate $A_2$ one can use the standard in-in analysis involving  the typical 
interaction ${\cal L}_3$ given above. Here we present an alternative derivation of $A_2$ as follows. Since $p_1 \ll q$, we can use the long-short decomposition as employed in Ref. \cite{Maldacena:2002vr} and write
\ba
\big \langle    \delta \phi_{\bfp_1}  \delta \phi_\bfq \delta \phi_{-\bfp_1- \bfq}
\big\rangle 
&\simeq& \left \langle  \delta \phi_{\bfp_1}  \Big(  \frac{d}{d  \phi} \left\langle  \delta \phi_\bfq \delta \phi_{- \bfq}\right\rangle\big|_{t_*} \delta \phi_{-\bfp_1}  \Big)\right \rangle \nonumber\\
&=& \big \langle  \delta \phi_{\bfp_1} \delta \phi_{-\bfp_1} \big \rangle \frac{d}{d \phi} \left\langle  \delta \phi_\bfq \delta \phi_{- \bfq}\right\rangle \big|_{t_*}  \, ,
\ea
in which $t_*$ represents the time in which the small scale mode $q$ leaves the horizon during the USR phase. Since $\delta \phi$ is frozen on superhorizon scale 
we can relate $\delta \phi$ at the time of end of inflation (or at the end of attractor phase in the final SR regime) to its value at horizon crossing. 
On the other hand, on a flat hypersurface $\delta \phi = - \sqrt{2 \epsilon}\calR= H \sqrt{2 \epsilon} d t $ so we can trade $d \phi$ by $d t$ and obtain
\ba
\big \langle    \delta \phi_{\bfp_1}  \delta \phi_\bfq \delta \phi_{-\bfp_1- \bfq}
\big\rangle  &\simeq&  \big \langle  \delta \phi_{\bfp_1} \delta \phi_{-\bfp_1} \big \rangle \frac{1}{\sqrt{2 \epsilon(t_*) } a(t_*) H} \frac{d}{d \tau} \left\langle  \delta \phi_\bfq \delta \phi_{- \bfq}\right\rangle \big|_{\tau_*} \, ,
\ea
in which $\tau$ is the conformal time with $d \tau = dt/a(t)$. 
On the other hand, to good accuracy 
\ba
\delta \phi(\tau)  =\frac{H}{\sqrt{2 q^3}} (1+ i q \tau) e^{-i q \tau} \, ,
\ea 
and hence 
\ba
 \frac{d}{d \tau} \left\langle  \delta \phi_\bfq \delta \phi_{- \bfq}\right\rangle = \frac{H^2 \tau}{q} \, .
\ea
Now, using the relations $q \tau_* \simeq -1$ and $\epsilon(\tau) = \epsilon_e\big(\tau/\tau_e \big)^6$ during the USR phase, we obtain
\ba
A_2 \simeq  \frac{N'' N' H^2 \tau_e^3}{\sqrt{2 \epsilon_e}}     \big \langle  \delta \phi_{\bfp_1} \delta \phi_{-\bfp_1} \big \rangle 
 \int \frac{d^3 q}{(2 \pi)^3}  \sim 
\frac{N'' N' H^2 }{3 \sqrt{2 \epsilon_e}}  \big \langle  \delta \phi_{\bfp_1} \delta \phi_{-\bfp_1} \big \rangle \, ,
\ea
where the integration is limited to modes which leave the horizon during the USR period, $q_s \leq q \leq q_e$. 
Using the expression for power spectrum given in Eq. (\ref{PR}), we finally obtain
\ba
\label{loop-A2}
\calP_\calR^{\mathrm{loop}}(\bfp)|_{A_2} \sim \eta_s e^{- 3 |N_s|}  
P_\calR(\bfp) \, \calP_\calR^{\mathrm{short}} \, ,
\ea
in which $\calP_\calR^{\mathrm{short}} \equiv (H^2/8 \pi^2 \epsilon_e)$ is the power spectrum associated to short modes which leave the horizon during the USR phase.
Using a direct in-in analysis with the action (\ref{L3}) (see  Ref. \cite{Cai:2018dkf} for details about the action ${\cal L}_3$)  one obtains a  result similar to Eq. (\ref{loop-A2}).  We see that  the  loop corrections induced from intrinsic non-Gaussianity is too small to be dangerous.  More specifically, compared to Refs. \cite{Kristiano:2022maq,  yokoyamanew} we have the  suppression factor 
$\eta_s e^{- 3 |N_s|} $ which make the loop contribution (\ref{loop-A2}) harmless.

Finally, the term $A_3$ is given by
\ba
\label{A3}
A_3 \equiv N' N'''    |\delta \phi_{\bfp} |^2
\int \frac{d^3 \bfq}{( 2 \pi)^3}   |\delta \phi(\bfq, \tau=0)|^2 = \frac{N'''}{N'} P_\calR(\bfp) \int \frac{d^3 \bfq}{( 2 \pi)^3}  |\delta \phi(\bfq, \tau=0)|^2 \, .
\ea
To calculate the integral above, we note that $\calR$ is smooth across the transitions while $\delta \phi(\bfq)$ is not. This is because $\dot \epsilon$ undergoes jumps at the start and end points of USR phase. Using the analysis of  Ref. \cite{f}, the curvature perturbation in the final phase is given by  
\ba
\calR^{(3)}_{k} =  \frac{H}{ M_P\sqrt{4 \epsilon(\tau) k^3}}  
\Big[ \alpha^{(3)}_k ( 1+ i k \tau) e^{- i k \tau}  + \beta^{(3)}_k ( 1- i k \tau) e^{ i k \tau}  \Big] \, ,
\ea
with the coefficients $\alpha^{(3)}_k$ and $\beta^{(3)}_k$ are given by,
\ba
\label{alpha-beta3}
\alpha^{(3)}_k = \frac{1}{8 k^6 \tau_i^3 \tau_e^3}  \Big[ 3h
 ( 1 -i k \tau_e)^2 (1+i k \tau_i)^2 e^{2i k (\tau_e- \tau_i)}
-i (2 k^3 \tau_i^3 + 3i k^2 \tau_i^2 + 3 i) (4 i k^3 \tau_e^3- h k^2 \tau_e^2 - h) \Big]
\nonumber
\ea
and
\ba
\beta^{(3)}_k=   \frac{1}{8 k^6 \tau_i^3 \tau_e^3}  \Big[ 3 ( 1+ i k \tau_i)^2 ( h+ h k^2 \tau_e^2 + 4 i k^3 \tau_e^3 ) e^{-2 i k \tau_i} + i h ( 1+ i k \tau_e)^2  ( 3 i + 3 i k^2 \tau_i^2 + 2 k^3 \tau_i^3 ) e^{- 2 i k \tau_e}\, .
 \Big] \nonumber
\ea 
Using  $\calR = -(H/\dot \phi) \delta \phi$,  at the end of inflation we have 
 \ba
|\delta \phi(\bfq, \tau=0)|^2= \left(\frac{H}{2 \pi} \right)^2 | \alpha^{(3)}_q + \beta^{(3)}_q|^2\, .
\ea
Using the above values of $\alpha^{(3)}_q$ and $\beta^{(3)}_q$ and performing the integral and using the value of $N'''$ calculated in previous section we obtain
\ba
\calP_\calR^{\mathrm{loop}}(\bfp)|_{A_3} \simeq   \eta_s^2  |N_s| P_\calR(\bfp) \calP_\calR^{\mathrm{short}}  \, .
\ea
Again, this  contribution is suppressed by a factor $\eta_s^2$ compared to what we expect from large loop corrections such as obtained in  Ref. \cite{Kristiano:2022maq}. 

While we presented the specific forms of $A_1, A_2$ and $A_3$, one can proceed with the remaining terms in Eq. (\ref{power-total}). However, associated to higher orders of $A_n$, there will be $n$-th derivative of $N$ which makes the contribution of $A_n$ more SR suppressed compared to the first few terms. 

\section{IV. Conclusions}
\setcounter{section}{4}
\noindent
Using the $\delta N$ formalism, we have computed the correction to the large-scale power spectrum from the short modes which are amplified during the USR phase. Accounting carefully for the initial SR phase, we have shown that the loop corrections are negligible when the transition between the USR and the SR phase is smooth, as advocated in Refs. \cite{riotto1,riotto2}.


\newpage
\centerline{\bf Acknowledgements}
\vskip 0.2cm
\noindent
A.R. is supported by the
Boninchi Foundation for the project ``PBHs in the Era of
GW Astronomy". H. F. thanks Mohammad Hossein Namjoo for useful discussions.



\begin{thebibliography}{99}

 \bibitem{s1} P.~Ivanov, P.~Naselsky and I.~Novikov,
  Phys.\ Rev.\ D {\bf 50}, 7173 (1994).
  
  \bibitem{s2} J.~Garc\'{\i}a-Bellido, A.D.~Linde and D.~Wands,
  Phys.\ Rev.\ D {\bf 54} (1996) 6040
  \arXivold{astro-ph/9605094}.

  \bibitem{s3} 
  P.~Ivanov, Phys.\ Rev.\ D {\bf 57}, 7145 (1998)
  \arXivold{astro-ph/9708224}.
  
  
%
 \bibitem{deltaN1} D.~S.~Salopek and J.~R.~Bond,
Phys. Rev. D \textbf{42} (1990), 3936-3962.
 
 
 \bibitem{deltaN2}
M.~Sasaki and E.~D.~Stewart,
Prog. Theor. Phys. \textbf{95} (1996), 71-78
\arXivold{astro-ph/9507001}.

\bibitem{Abolhasani:2019cqw}
A.~A.~Abolhasani, H.~Firouzjahi, A.~Naruko and M.~Sasaki,
``Delta N Formalism in Cosmological Perturbation Theory,''
WSP, 2019,

\bibitem{deltaNloop} 
C.~T.~Byrnes, K.~Koyama, M.~Sasaki and D.~Wands,
JCAP \textbf{11} (2007), 027
\arXiv{0705.4096}{hep-th}.

\bibitem{Kristiano:2022maq}
J.~Kristiano and J.~Yokoyama,
\arXiv{2211.03395}{hep-th}.



\bibitem{riotto1} A.~Riotto,
\arXiv{2301.00599}{astro-ph.CO}.

\bibitem{yokoyamanew}
J.~Kristiano and J.~Yokoyama,
 \arXiv{2303.00341}{hep-th}.
 

\bibitem{riotto2} A.~Riotto,
\arXiv{2303.01727}{astro-ph.CO}.

\bibitem{Choudhury:2023vuj}
S.~Choudhury, M.~R.~Gangopadhyay and M.~Sami,
\arXiv{2301.10000}{astro-ph.CO}.

\bibitem{f} H.~Firouzjahi,
\arXiv{2303.12025}{astro-ph.CO}.

\bibitem{m} H.~Motohashi and Y.~Tada,
\arXiv{2303.16035}{astro-ph.CO}.




\bibitem{Namjoo:2012aa} 
  M.~H.~Namjoo, H.~Firouzjahi and M.~Sasaki,
  Europhys.\ Lett.\  {\bf 101}, 39001 (2013)
  \arXiv{1301.5699}{hep-th}.
  
\bibitem{Chen:2013eea}
X.~Chen, H.~Firouzjahi, E.~Komatsu, M.~H.~Namjoo and M.~Sasaki,
JCAP \textbf{12}, 039 (2013)
\arXiv{1308.5341}{astro-ph.CO}.
  
\bibitem{Cai:2018dkf}
Y.~F.~Cai, X.~Chen, M.~H.~Namjoo, M.~Sasaki, D.~G.~Wang and Z.~Wang,
JCAP \textbf{05}, 012 (2018)
\arXiv{1712.09998}{astro-ph.CO}.
  
\bibitem{Maldacena:2002vr} 
  J.~M.~Maldacena,
  JHEP {\bf 0305}, 013 (2003)
\arXivold{astro-ph/0210603}.


%
 
%
%
%
%
%
%
%


    
    
    
%
%
%
%
%
%


%
%
%

%
%
%
%
%
%
%
%
%
%
%
%
%
%
%
%
%



\end{thebibliography}
\end{document}